\documentclass[sn-mathphys]{sn-jnl}% Math and Physical Sciences Reference Style
\jyear{2021}%

\theoremstyle{thmstyleone}%
\theoremstyle{thmstyletwo}%

\theoremstyle{thmstylethree}%

\begin{document}

\title[]{Efficient Coding Approach Towards Non-Linear Spectro-Temporal Receptive Fields}


\author*[1]{\fnm{Pranav} \sur{Sankhe}}\email{pranavsankhe40@gmail.com}

\author[2]{\fnm{Prasanna} \sur{Chaporkar}}


\affil[1]{\orgdiv{Center for Information and Neural Networks}, \orgaddress{\state{Osaka}, \country{Japan}}}

\affil[2]{\orgdiv{Electrical Engineering Department}, \orgname{IIT Bombay}, \orgaddress{\city{Mumbai}, \country{India}}}

% \affil[3]{\orgdiv{Department}, \orgname{Organization}, \orgaddress{\street{Street}, \city{City}, \postcode{610101}, \state{State}, \country{Country}}}


\abstract{Linear Non-Linear(LN) models are widely used to characterize the receptive fields of early-stage auditory processing. We apply the principle of efficient coding to the LN model of Spectro-Temporal Receptive Fields(STRFs) of the neurons in primary auditory cortex. The Efficient Coding Principle has been previously used to understand early visual receptive fields and linear STRFs in auditory processing. Efficient coding is realized by jointly optimizing the mutual information between stimuli and neural responses subjected to the metabolic cost of firing spikes. We compare the predictions of the efficient coding principle with the physiological observations, which match qualitatively under realistic conditions of noise in stimuli and the spike generation process.}




\maketitle
Information theoretical concepts like entropy and mutual information are useful measures to describe the coding efficiency of sensory systems. Applications of the efficient coding principle to early visual and linear auditory systems have helped explain the characteristics of neuronal receptive fields (Srinivasan et al., 1982; Atick et al., 1990; Lewicki, 2002; Smith \& Lewicki, 2006). In particular, neural responses have been found to be optimized to transmit as much information as possible about the naturally occurring stimuli. Furthermore, this optimization is subjected to the cost of generating the neural response (Laughlin et al., 1998; Lennie, 2003). For instance, early visual representations in the retina, LGN, and primary visual cortex have been shown to be consistent with the principles of information maximization in the presence of noise. In the auditory system, the spectro-temporal receptive fields (STRFs) have also been shown to maximize the information received from the stimulus along with minimizing the cost of generating spikes. (Lewicki, 2002; Smith Lewicki, 2006)\\

The efficient coding principle has been used to explain the dependency of STRFs on the stimulus properties (Lesica \& Grothe, 2006; Zhao, 2011). In addition to that, multiple experimental studies have shown throughout auditory code that includes the cochlea, the auditory nerve, and the auditory cortex follow optimal coding (Attias et al., 1997; Rieke et al. 1995; Lewicki, 2002;). Interestingly enough, evidence for the efficient coding hypothesis has also been demonstrated by behavioural experiments on human auditory perception (Ming et al., 2009).\\

Traditional STRFs are linear kernels that characterize the spectral and temporal selectivity of a neuron to the incoming stimulus. More specifically, STRF explains the sensitivity of a neuron at a particular latency $t$ to acoustic frequency $f$ by relating the spike rate of the neuron with the time-frequency representation (spectrogram) of the stimulus. However, along the auditory pathway, there are several nonlinearities in the neural responses which the linear STRFs fail to capture. For example, the neurons in auditory systems of primates, bats and birds (Rauschecker et al., 1995; Ohlemiller et al., 1996; Scheich et al., 1979) are tuned explicitly to only complete animal vocalizations. Additionally, auditory neurons also have been found to be selective to the amplitude of sounds in the same frequency range (Nelken et al., 2008) and to sounds of varying levels of spectral contrast (Barbour et al., 2003). These nonlinearities in auditory coding have inspired several nonlinear models of STRFs, such as the Linear-Nonlinear (LN) model. In this paper, we focus on applying the efficient coding principle to a non-linear model of STRF of neurons in the primary auditory cortex.\\

The LN model, which is an instantiation of a generalized linear model (GLM), is used to characterize the early stages of auditory processing (Aertsen \& Johannesma, 1981; Radtke-Schuller et al., 2020). LN STRFs have been used successfully to model the receptive fields of the auditory cortex, and it has been shown that compared to the linear model of STRF, the LN model estimates the neural responses significantly better (Machens et al., 2004; Simoncelli et al., 2004). LN model is conceptualized by applying a static non-linearity onto the linear approximation of the spike generating process. In the linear stage of the model, a FIR filter $h(f,\tau)$ is applied to the time-frequency representation (spectrogram) of the stimulus ($s(f,t)$) to produce spike rate $r(t)$.
\begin{equation}
r(t) = \sum_{f=q}^{f=F}\sum_{\tau=0}^{\tau=T}h(f,\tau)s(f, t-\tau) \label{1}
\end{equation}
$h(f, \tau)$ is often referred to as the linear Spectro-Temporal Receptive Field (STRF) in the auditory system. Let $\boldsymbol{r}$ be the output vector of $N$ neurons representing the firing rate. In the case of populations of neurons, let $\boldsymbol{S}$ be the input time-frequency representation of the audio stimulus and $\boldsymbol{h_i}$ be the linear filter of neuron $i$. The response of neuron $i$ is computed by taking an inner product of noise corrupted input and linear filter to obtain membrane voltage $y_i$. This voltage is passed through a static nonlinearity to produce the spikes, which are corrupted with output neural noise. The current study used a rectified linear function (Agarap, 2018) for the output non-linearity in the second stage of the LN model. Each neuron is characterized with an offset ($b_i$) and slope($g_i$) for its rectified linear function. The neural response in terms of firing rate is directly calculated by applying the RELU function on the membrane voltage ($y_i$).
\begin{equation}
y_i = \boldsymbol{h_i}(\boldsymbol{S} + \boldsymbol{n_s}) \label{2}    
\end{equation}
\begin{equation}
r_i = g_i(y_i + b_i) + n_r \label{3}    
\end{equation}
% 
% 
% 
% \subsection{Optimization Problem}\label{subsec2}
% 
We employ a generative linear-nonlinear model coupled with an efficient coding principle to optimize the linear filters and the static nonlinearities. The optimization problem is designed to maximize the mutual information between the natural sounds and the neural responses in the presence of input and output noise. The metabolic costs involved in generating the neural response impose constraints on neurons' firing rate, thus constraining mutual information maximization. We borrow techniques and intuitions from the work of Karklin and Simoncelli (Karklin \& Simoncelli, 2011) on receptive fields of retinal ganglion cells to derive and explain the results in this paper.\\

We consider the problem of jointly optimizing the linear filter and nonlinearities by maximizing the mutual information ($I(S, R)$) subject to neural costs such as metabolic energy costs involved in the spike generation process. Following the approach taken in vision, the neural costs are modelled to be proportional to the average spike rate of the neuronal population (Zhaoping, 2006). $\lambda_i$ parameterizes the trade-off between the cost of neural response and the information gained by generating the response. Mutual information can be expressed as the difference between average uncertainty about the identity of the stimulus given the response and total stimulus entropy, i.e., $H(S) - H(S \vert R)$. Since the global stimulus entropy $H(S)$ does not depend on the model, the optimization problem can be framed as maximizing the negative conditional entropy subject to the firing cost. Incorporating penalty on the firing rate, we get an objective function:
\begin{equation}
-H(S\vert R) - \sum_i{\lambda_i<r_i>} \label{4}
\end{equation}
The conditional entropy $H(S\vert R)$ is computed by calculating the expectation of covariance of the posterior $C_{s\vert r}$. $C_{s\vert r}$ is calculated by the covariance of the prior $C(r\vert s)$ multiplied by the linear filters $\boldsymbol{W}$ and the diagonal matrix of the nonlinear responses $\boldsymbol{G}$ and added to the inverse of the global covariance matrix of the input (Karklin \& Simoncelli, 2011). Using the identity that logarithm of determinant of a matrix is equal to the trace of the logarithm of the matrix, we the objective function simplifies to:
\begin{equation}
    -E[\frac{1}{2}log(2\pi e*trace(C_{s\vert r}))] - \sum_i\lambda_i<r_i> \label{5}
\end{equation}
The model is trained using across batches of data using gradient ascent on the objective function that encapsulates the coding efficiency. We use Tensorflow (Abadi et al., 2016) to perform the optimization with a learning rate of $0.001$ over $80,000$ iterations. We run the model for $36$ neurons. The verification of convergence of the LN model is not possible by simulating data, unlike the algorithms for other generative models like PCA or ICA. Cochleagrams of natural sounds are obtained from a large collection of human speech, ferret vocalizations, and recordings of the ambient laboratory environment (Lopez Espejo et al., 2019). Cochleogram is a spectral representation of the auditory nerve firing rate (Brown \& Cooke, 1994). Neural recordings from the primary auditory cortex of ferrets (Lopez Espejo et al., 2019), passively listening to the natural sounds and cochleagram of the stimulus were used to estimate the LN STRFs of auditory neurons. Spike rate data and the cochleagrams are fitted for each neuron using the gradient ascent algorithm to update the parameters. Preprocessing of the data and model estimation was done using the NEMS library in Python (David, 2018).\\ 

To investigate the role of input and output noise, we estimated STRFs from the model in high noise and low noise conditions. The high noise conditions were set to the ones present in the physiological data. The input and output noise was set to extremely low values (-40dB) in low noise conditions. We observed that in the presence of significant output noise, the STRFs generated by the model match with the estimated STRFs from the physiological data [Fig 1]. The model was able to estimate sharp rectifying non-linearities of individual neurons in the presence of significant output noise. On the other hand, in low noise conditions, the model was unable to learn the sharp rectification. The estimated slopes of almost all the neurons evaluate to zero. Past literature has suggested that hard rectification in neurons is more useful in the presence of noise (Carandini, 2004). The results support claims that the neurons in the early auditory regions optimally encode the incoming stimulus information.\\
\begin{figure}[!h]%
\centering
\includegraphics[width=0.9\textwidth]{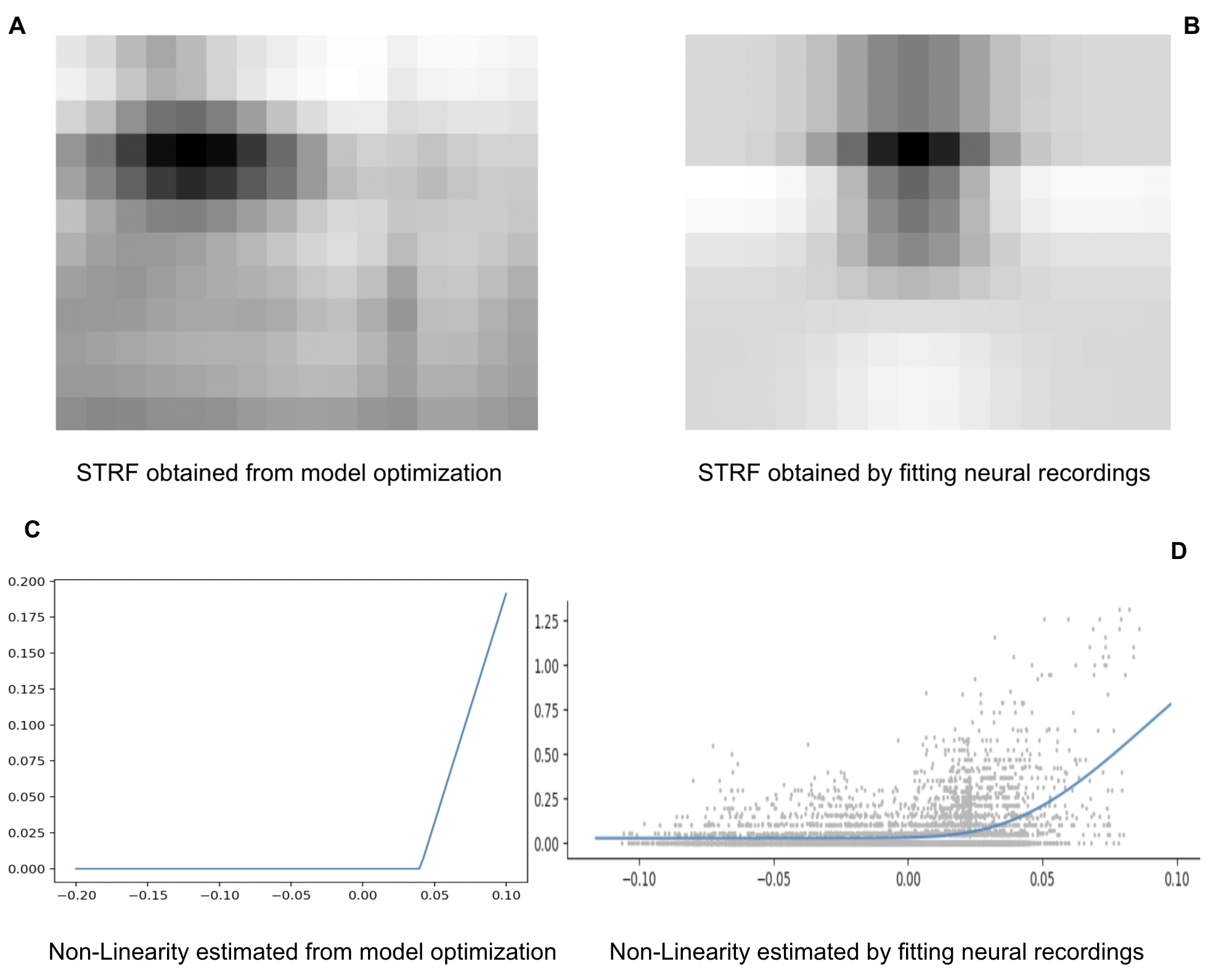}
\caption{Figure A is the LN STRF obtained by applying efficient coding principle whereas figure B shows the physiological LN STRF estimated from neural spike data recorded from the primary auditory cortex of one neuron of ferrets while listening to natural sounds. Figures C and D illustrate the estimated non-linearitiy and the measured non-linearity for a neuron respectively.}\label{fig1}
\end{figure}
Past work has mainly focused on the linear part of the STRF (Lesica \& Grothe, 2006; Zhao, 2011), and our results indicate that the efficient coding principle can be applied to non-linear models of STRF.
Here we show an efficient coding for non-linear STRFs under the presence of input and output noise. The linear filter of the STRF and static nonlinearities are estimated jointly for a population of neurons by maximizing the mutual information and between spike rates and auditory stimuli (cochleagrams). We demonstrate that the non-linear model STRFs derived using the efficient coding principle qualitatively matches with the physiological STRFs obtained by fitting neural data recorded from the primary auditory cortex of ferrets.

\end{document}